\documentclass[prd,aps,preprint]{revtex4}
\usepackage{graphicx}
\begin{document}
\tolerance=5000
\def\be{\begin{equation}}
\def\ee{\end{equation}}
\def\bea{\begin{eqnarray}}
\def\eea{\end{eqnarray}}
\def\ii{\'{\i}}
\def\bi{\bigskip}
\def\be{\begin{equation}}
\def\en{\end{equation}}
\def\bq{\begin{eqnarray}}
\def\eq{\end{eqnarray}}
\def\noi{\noindent}
\title{Acausality in Gowdy spacetimes}
\author{Hernando Quevedo}
\email{quevedo@nuclecu.unam.mx}
\affiliation{Instituto de Ciencias Nucleares\\
Universidad Nacional Aut\'onoma de M\'exico\\
A.P. 70-543,  M\'exico D.F. 04510, M\'exico\\}
\affiliation{Department of Physics \\
University of California \\
Davis, CA 95616}
\date{\today}
\begin{abstract}
\setlength{\baselineskip}{.5cm}
We present a parametrization of $T^3$ and $S^1\times S^2$ Gowdy
cosmological models which allows us to study both types of
topologies simultaneously. We show that there exists a coordinate
system in which the general solution of the linear polarized
special case (with both topologies) has exactly the same functional
dependence. This unified parametrization is used to investigate the
existence of Cauchy horizons at the cosmological singularities,
leading to a violation of the strong cosmic censorship conjecture.
Our results indicate that the only acausal spacetimes are described
by the Kantowski-Sachs and the Kerr-Gowdy metrics.
\end{abstract}
\maketitle


\section{INTRODUCTION}
\label{sec:int}

Gowdy cosmologies \cite{gowdy} have been studied for more than 30
years, especially motivated by the desire to understand the
mathematical and physical structure of singularities in
cosmological spacetimes (globally hyperbolic pseudo-Riemannian
manifolds with  compact Cauchy spatial hypersurfaces which satisfy
Einstein's field equations). It has been long suggested
\cite{clarke} that a singularity is characterized either by a blow
up of the curvature and tidal forces, or by a breakdown of
causality. However, it is not clear when either possibility is to
be expected. The singularity theorems \cite{hawkingellis} state
that reasonable matter evolves from regular data into a
singularity, when the evolution is governed by Einstein's
equations.  The singularities that form in such a process are
characterized by causal geodesic incompleteness. The generic
nature of these singularities, however, is not described by the
singularity theorems. In particular, the question about the blow
up of the curvature and tidal forces at the singularity and about
the existence of a horizon that ``hides'' the singularity cannot
be addressed with the methods used to study geodesic
incompleteness.

Cosmic censorship conjectures
state that a singularity must be hidden by an event horizon (weak
conjecture) or not be detectable by timelike observers until
they fall into it (strong conjecture). In the context of
the initial value problem of general relativity, the strong
cosmic censorship (SCC) asserts that in the space of allowed
initial data there exists only a very small set which evolves
into spacetimes that can be extended beyond their maximal
domain of dependence into acausal regions. Such extendible
spacetimes are characterized by the existence of Cauchy horizons.
When a Cauchy horizon exists in a given spacetime, one
expects that, in principle, timelike geodesics can be found
which become closed after crossing the Cauchy horizon.
This would indicate a violation of the SCC conjecture.

In the context of Gowdy cosmological models, the fundamental
questions concerning global existence of solutions \cite{mon1}
and the existence of cosmological singularities \cite{isemon}
were analyzed in detail for the case of a $T^3$ topology,
whereas the $S^1\times S^2$ case has received less attention.
In general, it has been shown that these spacetimes possess
cosmological singularities, and the asymptotic behavior of
the metric and curvature near these singularities has been the subject of
numerous studies (for a recent review, see \cite{ber1}).
The question of the curvature behavior at the cosmological
singularity has recently been answered in quite general terms in
\cite{chrulake} and \cite{ring1} for the case of a toroidal
topology. According to Isenberg and Moncrief \cite{isemon},
a Gowdy model is called generic when all the corresponding
curvature invariants blow up at the cosmological
singularity for all values of the spatial coordinate.
If the curvature invariants happen to remain regular
at the cosmological singularity, the spacetime is called
non-generic. In this case, the singularity could become
a Cauchy horizon and, in principle, it should be possible
to extend the spacetime across the Cauchy horizon to include
non-globally hyperbolic acausal regions, indicating a
violation of the SCC conjecture. Consequently, if we
want to maintain the predictability of Einstein's equations,
we should avoid the existence of Cauchy horizons.
It follows that within the class of Gowdy models, the
SCC conjecture can be violated only in non-generic models.
This paper is concerned with the search for non-generic
models with $T^3$ and $S^1 \times S^2$ topologies.
We will show that only a very small set (the Kerr-Gowdy metric)
of Gowdy models
are non-generic so that the SCC conjecture holds in most
models.

This paper is organized as follows. In Section \ref{sec:param} we
present a particular parametrization and a system of coordinates
in which the field equations exhibit the same functional dependence
for $T^3$ and $S^1\times S^2$ models.
This allows us to derive the general solution for Gowdy polarized
models
in Section \ref{sec:gensol}. In Section \ref{sec:cauchy} we
investigate the question of existence of Cauchy horizons
in the general solution. Finally,
in Section \ref{sec:conclusions}
we summarize our results.

\section{A unified parametrization of Gowdy cosmologies}
\label{sec:param}

Gowdy cosmological models are inhomogeneous time-dependent
solutions of Einstein's vacuum equations. The most general
topology of the spatial hypersurfaces can be shown to be
either $T^3$ or $S^1\times S^2$. In most
studies only the special case of polarized $T^3$ models has been
considered. Here we will use a particular
parametrization of the corresponding line element which
allows us to analyze both cases in quite general terms.
Let us introduce the line element
\be
ds^2 = e^{-\lambda/2 + \tau/2} (- e^{-2\tau} d\tau^2 + d\chi^2)
+ \sqrt{g_2} \left[e^P (d\sigma + Q d\delta)^2 + e^{-P} d\delta^2\right]\ ,
\label{gle}
\ee
where $P$, $Q$, and $\lambda$ depend on the non-ignorable coordinates
$\tau$ and $\chi$. The cosmological models are compactified by
requiring that  $0\leq \chi, \sigma, \delta \leq 2\pi$. The function
$g_2$ corresponds to the determinant of a two-metric and satisfies
the differential equation $(t=e^{-\tau})$
\be
\frac{g_{2,tt}}{g_2} - \frac{1}{2}\left(\frac{g_{2,t}}{g_2}\right)^2
-
\frac{g_{2,\chi\chi}}{g_2} + \frac{1}{2}\left(\frac{g_{2,\chi}}{g_2}\right)^2 = 0
\label{eqg2}
\ee
which follows from the vacuum field equations. The special case
of a $T^3$ topology is obtained from the solution
\be
g_2 = t^2 = e^{-2\tau}\ ,
\label{gst3}
\ee
while the $S^1\times S^2$ case corresponds to
\be
g_2 = c^2 \sin^2 e^{-\tau} \sin^2\chi \ ,
\label{g2s1s2}
\ee
where $c$ is a real constant.
It turns out that the field equations reduce to a set of two second-order
coupled partial differential equations for $P$ and $Q$ and a set of two
first-order partial differential equations for $\lambda$ which can be
integrated by quadratures once $P$ and $Q$ are known.
In the following analysis we will consider only the main field equations
for $P$ and $Q$.
To handle these equations it is appropriate to consider the
corresponding Einstein-Hilbert Lagrangian ${\cal L} = \sqrt{-g}R$
which after a Legendre transformation can be written as \cite{sigma}
\be
{\cal L} = \frac{1}{2} e^\tau \sqrt{g_2} [P_{,\tau}^2 - e^{-2\tau} P_{,\chi}^2
+ e^{2P} (Q_{,\tau}^2 - e^{-2\tau} Q_{,\chi}^2)] \ .
\label{ehl}
\ee
This corresponds also to the Lagrangian of a non-linear sigma model
$SL(2,R)/SO(2)$ where the target space turns out to be a hyperbolic space
with metric $ds_2^2 = dP^2 + e^{2P} d Q^2$, when $Q$ and $P$ are used
as coordinates of the target space \cite{gowdy}. Accordingly,
the Gowdy cosmologies can be considered in general as a special case of the
non-linear sigma model $SL(2,R)/SO(2)$. The variation of the Lagrangian
(\ref{ehl}) yields
\be
P_{,\tau \tau}-e^{-2 \tau} P_{,\chi \chi}+
P_{,\tau}(1+g_{2}^{-1}g_{2, \tau})-e^{-2 \tau}g_{2}^{-1}g_{2, \chi}
P_{,\chi}-e^{-2 P}(Q_{,\tau}^{2}-e^{-2 \tau }Q_{,\chi}^{2})=0\ ,
\label{eqp}
\ee
\be
Q_{,\tau \tau}-e^{-2 \tau} Q_{,\chi \chi}+Q_{\tau}(1+g_{2}^{-1}g_{2, \tau})
-e^{-2 \tau}g_{2}^{-1}g_{2, \chi}Q_{,\chi}+2(P_{,\tau}Q_{,\tau}-e^{-2 \tau}
P_{,\chi}Q_{,\chi})=0\ .
\label{eqq}
\ee
The important aspect about the Lagrangian (\ref{ehl}) is that it
can be used to derive a more compact representation of the main field
equations. This is the so called Ernst representation \cite{ernst}
which was originally derived for axisymmetric stationary spacetimes
and has been generalized to include different types of spacetimes
with two Killing vector fields \cite{nora}. In the case of Gowdy
cosmologies (\ref{gle}), the Ernst equation can be written as \cite{prd}
\be
(1-\xi\xi^*)\left[ \nabla^2\xi + \frac{1}{2}\nabla\ln(g_2)\ \nabla \xi\right]
+ 2\xi^*(\nabla \xi)^2 = 0 \ ,
\label{ernsteq}
\ee
where $\nabla = (\partial_t,\, i\partial_\chi)$ is a complex vector operator,
$t = e^{-\tau}$, and the Ernst potential is defined as
\be
\xi = \frac{1-\sqrt{g_2} \ e^P - i R}{1+\sqrt{g_2} \ e^P + i  R} \ ,
\qquad
R_{,t} = \sqrt{g_2} e^{2P} \, Q_{,\chi} \ ,
\quad
R_{,\chi} = \sqrt{g_2} e^{2P} \, Q_{,t} \ .
\label{ernstpot}
\ee
Here an asterisk represents complex conjugation. By using Eq.(\ref{eqg2}),
it is straightforward to
show that in fact the Ernst equation (\ref{ernsteq}) is equivalent to
the main field equations (\ref{eqp}) and (\ref{eqq}). As we can see from
the above expressions, in this parametrization the only difference
between $T^3$ and $S^1\times S^2$ models lies in the determinant $g_2$.
This difference can be ``hidden" if we success in finding a representation
in which the determinant coincides for both cases. This can easily be achieved
by introducing coordinates $x$ and $y$  for the $T^3$ case as
\be
e^{-2\tau} = c^2 (1-x^2)(1-y^2) \ , \qquad \chi = cxy \ ,
\label{xyt3}
\ee
with $x^2\leq 1$ and $y^2\leq 1$,
and the coordinates $\tilde x$ and $\tilde y$ for the $S^1\times S^2$ topology as
\be
\tilde x = \cos e^{-\tau} \ , \qquad \tilde y = \cos\chi \ ,
\label{xys1s2}
\ee
so that the determinant becomes $g_2 = c^2(1-x^2)(1-y^2)$ for the $T^3$ case, while
for the $S^1\times S^2$ case we get the same expression with $x$ and $y$
replaced by $\tilde x$ and $\tilde y$, respectively. In these coordinates,
the Ernst equation (\ref{ernsteq}) can be written as
\be
(1-\xi\xi^*)\{ [(1-x_*^2)\xi_{,x_*}]_{,x_*} - [(1-y_*^2)\xi_{,y_*}]_{,y_*}\}
+2\xi^* [ (1-x_*^2)\xi_{,x_*}^2 - (1-y_*^2)\xi_{,y_*}^2] = 0 \ ,
\label{xixy}
\ee
where $x_* = x$, $y_*= y$ for $T^3$ models, and $x_* = \tilde x$, $y_*= \tilde y$
for $S^1\times S^2$ models. Thus, we
have obtained a representation in which the main field equations for all
Gowdy cosmologies have the same functional dependence.
For the sake of completeness, we also present the final form of the
general line
element in the new coordinates:
\be
ds^2 = e^{-\lambda_*/2}\left( -\frac{dx_*^2}{1-x_*^2} + \frac{dy_*^2}{1-y_*^2}\right)
+ c(1-x_*^2)^{1/2}(1-y_*^2)^{1/2}[e^P(d\sigma + Q d\delta)^2 + e^{-P} d\delta^2]
\ ,\label{glexy}
\ee
where
\be
e^{-\lambda_*/2} =
c^{3/2}  \frac{x^2-y^2}{ (1-x^2)^{1/4}(1-y^2)^{1/4} }e^{-\lambda/2} \ ,
\quad
e^{-\lambda_*/2} =
(\arccos \tilde x)^{-1/2}e^{-\lambda/2} \ ,
\ee
for the $T^3$ and $S^1\times S^2$ models, respectively.
In this parametrization, the only
functional difference
between both topologies is contained in the form
of the metric function $\lambda_*$.

\section{The general solution}
\label{sec:gensol}

The importance of the parametrization of the last section is that it
allows us to investigate both types of Gowdy models with the same
functional dependence. Let us consider now the special polarized
case, $Q=0$. From the definition of the Ernst potential
(\ref{ernstpot}) we see that the function $R$ reduces to a constant
which, without loss of generality, can be put as $R=0$.
A straightforward calculation shows that in this case the Ernst
equation (\ref{ernsteq}) reduces to
\be
[(1-x_*^2) P_{,x_*}]_{,x_*} - [(1-y_*^2) P_{,y_*}]_{,y_*} = 0 \ ,
\label{eqpxy}
\ee
an equation which can be solved by separation of
variables and whose general solution can be written as
an infinite series
\cite{erdely}
\be
P= \sum_{\nu} [a_\nu P_\nu(x_*) + b_\nu Q_\nu(x_*)]
[c_\nu P_\nu(y_*) + d_\nu Q_\nu(y_*)] \ ,
\label{gensol}
\ee
where $\nu$ is a constant, $P_\nu$ and $Q_\nu$ are
the Legendre functions of first and second kind, respectively,
and $a_\nu$, $b_\nu$, $c_\nu$ and $d_\nu$ are real constants.

It is now a question of analyzing the behavior of the functions
$P_\nu$ and $Q_\nu$ within the interval $-1 \leq x_*,\, y_*
\leq +1$ to determine which of the solutions contained
in (\ref{gensol}) are physical relevant. For instance, one should
impose that the function $P$ is periodic in the angular coordinate
$\chi$. This condition is identically satisfied in the $S^1\times S^2$ case
because the angular dependence of the general solution is determined
through  $y_* = \tilde y = \cos\chi$. In the $T^3$ case one can also
show \cite{erdely} that (\ref{gensol}) contains an infinite
 number of periodic solutions.
Furthermore, it is possible to analyze the asymptotic behavior
of the solution in quite general terms. If $\nu$ is not an integer,
the function $P_\nu$ diverges at $x_*=-1$. But if $\nu$
is an integer number, say $n$, then $P_\nu$ becomes the Legendre
polynomials $P_n$ which are free of singularities for any values in
the interval $-1\leq x_*\leq +1$. On the other hand, the function
$Q_\nu$ possesses singularities at $x_*=\pm 1$ for all integer
and non-integer values of $\nu$.

An additional important aspect of the solution presented above is
that it coincides exactly with the general static axisymmetric
solution of Einstein's vacuum equations \cite{erro}  in prolate
spheroidal coordinates. Moreover, the Ernst equation
(\ref{ernsteq}) is functionally equivalent to the main field
equations of stationary axisymmetric spacetimes. This functional
equality is due to  the fact that both Gowdy cosmologies and
stationary axisymmetric spacetimes possess a set of two commuting
Killing vector fields. The Ernst equation has been used
to analyze the internal symmetries of the field equations and to
develop the modern solution generating techniques. In particular,
in a recent work \cite{jmp} it was shown that all the Gowdy $T^3$
cosmologies can be generated from the data at the initial
singularity. The results presented here suggest that a
similar procedure can be developed for $S^1\times S^2$ models. Our
results also explain why the Kantowski-Sachs metric (the region
inside the horizon of the Schwarzschild metric) and the Kerr-Gowdy
metric (the region inside the horizons of the Kerr metric) can be
interpreted both as $T^3$ \cite{jmp} and $S^1\times S^2$ Gowdy
cosmological models \cite{prd}.

\section{Cauchy horizons}
\label{sec:cauchy}

In the previous section we have derived a unified parametrization
for all types of Gowdy cosmologies and found the general polarized
solution. In this section we will show that this general solution
contains all the information necessary to determine which spacetimes
can allow the existence of Cauchy horizons. As we mentioned in
the Introduction, a cosmological singularity can become a Cauchy
horizon if the curvature is regular at the singularity.

The cosmological singularities of $S^1\times S^2$
Gowdy models in the original parametrization $(\tau, \ \chi)$
correspond to the limits $\tau \rightarrow \infty$
and $\tau\rightarrow -\ln\pi$. In the
coordinates $\tilde x, \ \tilde y$ described
above, this corresponds to the hypersurfaces $\tilde x \rightarrow  1$
and $\tilde x \rightarrow  -1$, respectively.
In the case of $T^3$ models, the singularity is situated at
$\tau \rightarrow \infty$, a limit that in coordinates $x$ and $y$
corresponds to $x^2 \rightarrow 1$ or $y^2 \rightarrow 1$.
According to the explicit form of the general line element
(\ref{glexy}), the latter case corresponds to a spatial
limit which is not of interest for the study of
cosmological singularities (temporal limit).
Therefore we can eliminate all possible singularities
at $y_*^2=1$ from the general solution (\ref{gensol}).
To avoid the singularity of the function
$P_\nu(y_*)$ at $y_*=-1$, we consider only positive
integer values of the constant $\nu$, i. e.,
$\nu = n = 0,1,2,...$. Furthermore, the singularities of the function
$Q_n(y_*)$ at $y_*=\pm 1$ can be eliminated by choosing
$d_n = 0$ in (\ref{gensol}).
Then the general solution reduces to
\be
P= \sum_{n} [a_n P_n(x_*) + b_n Q_n(x_*)]
c_n P_n(y_*)  \ .
\label{gensol1}
\ee
We now consider the singularity at $x_*=\pm 1$.
The Legendre polynomials $P_n(x_*)$ and their derivatives
have constant regular values at the limits $x_*=\pm 1$.
So they essentially do not contribute to the behavior
of the solution at the cosmological singularity, and
we can completely ignore its contribution by choosing $a_n=0$.
Consequently, the general solution which is of importance
at the singularity can be written as
\be
P= \sum_{n}  b_n  Q_n(x_*) P_n(y_*)  \ ,
\label{gensol2}
\ee
where we have chosen $c_n=1$, without loss of generality.
The main point now is that the functional dependence of
the cosmological solution (\ref{gensol2}) coincides
exactly with the functional dependence of static
axisymmetric asymptotically flat solutions. The only
difference lies in the physical meaning of the coordinates
$x_*$ and $y_*$. While for static solutions both coordinates
are spacelike, in polarized Gowdy cosmologies the coordinate $x_*$
becomes timelike. The condition of asymptotic flatness,
which is used to obtain the general solution for static
spacetimes in the form (\ref{gensol2}), corresponds in
polarized Gowdy spacetimes to the condition of considering only
those solutions which are non-ignorable for the analysis
of the asymptotic behavior near the cosmological
singularities. Moreover, the behavior of the general
solution (\ref{gensol2}) near the cosmological singularity
corresponds to the near horizon limit of static solutions
$x_*\to 1$ (in spherical coordinates, this corresponds to
the limit $r\to 2m$, where $m$ is the Schwarzschild mass).

Let us now consider the unpolarized case, $Q\neq 0$.
The static counterpart of the general polarized solution
(\ref{gensol2}) has been used
to derive the most general stationary axisymmetric
asymptotically flat spacetime by using solution generating
techniques \cite{prd86}.
This generalized solution contains, in particular,
the Kerr metric which is the most general (vacuum)
black hole solution. In fact, the uniqueness theorems
\cite{heusler}
state that the Kerr spacetime is the most general
solution with {\it regular} horizons.
On the other hand, the inner and outer horizon limits
of the Kerr metric correspond to $x_* \to
-1,\ +1$, respectively \cite{prd}.
From the above considerations,
it is clear that the general polarized ($Q=0)$  solution
(\ref{gensol2}) can be used to generate the most
general unpolarized ($Q\neq 0$) Gowdy solution which
should be considered for analyzing the behavior near
the cosmological singularities. In particular,
the Kerr-Gowdy solution must be contained
as a special case. For this solution, it has been
shown \cite{prd} that the cosmological singularities
are situated at $x_* \to \pm 1$, a limit that coincides
with the near horizon limit of the Kerr metric.
Using the functional analogy between stationary spacetimes
and unpolarized Gowdy cosmologies and
the black hole uniqueness theorems, we can conclude
that the Kerr-Gowdy metric is the most general
solution with a regular curvature behavior near
the cosmological singularities, i.e. it is the
most general non-generic Gowdy spacetime.
 The hypersurfaces
$x_*=\pm 1$ could become Cauchy horizons so that
the Kerr-Gowdy spacetime could be extended to
include acausality regions where the SCC conjecture
is violated. In fact, in a recent work \cite{zet}
several generalizations of the Kantowski-Sachs
and the Kerr-Gowdy spacetimes
have been analyzed, finding in all of them a curvature blow up
at the cosmological singularities which
does not allow the formation of Cauchy horizons.

\section{CONCLUSIONS}
\label{sec:conclusions}

In this work we have found a unified parametrization
for $T^3$ and $S^1\times S^2$ Gowdy cosmological models.
This unified parametrization allowed us to find the general
polarized solution for both types of Gowdy models in terms
of Legendre functions of first and second kind. We analyzed
the general solution which determines the behavior
near the cosmological singularities. Using the functional
analogy between Gowdy cosmologies and stationary axisymmetric
spacetimes, we concluded that the Kerr-Gowdy metric is the
most general Gowdy cosmological model in which acausality
regions might exist that violate the SCC conjecture.
This result agrees with the conclusion of \cite{isemon}
that only a very small set of polarized Gowdy spacetimes could be
extended into an acausal region, across
a Cauchy horizon. In fact, we have shown that this
set includes only the Kantowski-Sachs spacetime. For
unpolarized models, our results show that only the Kerr-Gowdy
metric could possess Cauchy horizons.
It would be interesting to show explicitly the existence
of closed timelike curves beyond the Cauchy horizons
of the Kerr-Gowdy metric. In the special polarized case of
the Kantowski-Sachs metric a preliminary study \cite{samos}
seems to indicate that this is impossible because for all
timelike geodesics there exists a focusing point which
does not allow them to cross the horizon. A more detailed
analytical study is necessary in order to completely
clarify this question.

\section*{ACKNOWLEDGMENTS}

In 1999, Mike Ryan presented a seminar at UNAM about the problem of
finding exact solutions which would describe the gravitational field
in Gowdy cosmologies. This was the starting point of a joint
collaboration, together with Octavio Obreg\'on, in which we are exploring
several methods of generating exact Gowdy cosmologies and are, simultaneously,
 reinterpreting
stationary spacetimes as Gowdy cosmologies. I would like to thank
Mike for drawing me closer to the world of the initial singularity.
It is a great pleasure to dedicate this work to my friend and colleague
Mike Ryan on the occasion of his sixtieth birthday.

This work was supported by US DOE grant DE-FG03-91ER 40674,
DGAPA-UNAM grant IN112401, CONACyT-Mexico grant 36581-E, and
UC MEXUS-CONACyT (Sabbatical Fellowship Program).


\begin{thebibliography}{99}

\bibitem{gowdy}  R. Gowdy, Phys. Rev. Lett. {\bf 27}, 826 (1971);
Ann. Phys. (N.Y.) {\bf 83}, 203 (1974).

\bibitem{clarke} C. J. S. Clarke, J. Math. Anal. Appl. {\bf 88}, 270
(1982).

\bibitem{hawkingellis} S. W. Hawking and G. F. R. Ellis, {\it The large scale
structure of spacetime} (Cambridge University Press, Cambridge, UK, 1973).

\bibitem{mon1} V. Moncrief, Ann. Phys., NY {\bf 132}, 87 (1981).

\bibitem{isemon} J. Isenberg and V. Moncrief, Ann. Phys. (N.Y.) {\bf 199}, 84 (1990).

\bibitem{ber1} B. Berger, Liv. Rev. Rel. {\bf 5}, 1 (2002).

\bibitem{chrulake} P. T. Chru\'sciel and K. Lake, Class. Quantum Grav.
{\bf 21}, S153 (2004).

\bibitem{ring1} H. Ringstr\"om, Class. Quantum Grav.
{\bf 21}, S305 (2004).


\bibitem{sigma} J. Cortez, D. N\'u\~nez, and H. Quevedo, Int. J. Theor. Phys.
 {\bf 40}, 251 (2001).

\bibitem{ernst} E. J. Ernst, Phys.Rev.  {\bf D 168}, 1415 (1968).

\bibitem{nora} J. B. Griffiths, {\it Colliding plane waves in general
relativity} (Oxford University Press, Oxford, 1991);
N. Breton, A. Feinstein and J. Iba\~nez, Class. Quantum
Grav. {\bf 9}, 2437 (1992);
N. Breton, Gen. Rel. Grav. {\bf 25}, 567 (1993).

\bibitem{prd}
O. Obregon, H. Quevedo, and M. P. Ryan, Phys. Rev. D {\bf 65}, 024022 (2001).

\bibitem{erdely} A. Erdelyi, W. Magnus, F. Oberhettinger, and F. G. Tricomi,
{\it Higher transcendental functions} (University of Chicago Press, Chicago, 1953).

\bibitem{erro}
G. Erez and N. Rosen, Bull. Res. Council of Israel
{\bf 8F}, 47 (1959);
C. Reina and A. Treves, Gen. Rel. Grav. {\bf 7}, 817 (1976).

\bibitem{jmp} A. Sanchez, A. Macias, and H. Quevedo, J. Math. Phys. {\bf 45},
1849 (2004).

\bibitem{kramernew} H. Stephani, D. Kramer, M. MacCallum, C. Hoenselaers,
and E. Herlt,
{\it Exact solutions of Einstein's field equations} (Cambridge University Press,
Cambridge UK, 2003).

\bibitem{prd86} H. Quevedo, Phys. Rev. D {\bf 33}, 324 (1986);
Fortschr. Phys. {\bf 38}, 733 (1990).


\bibitem{heusler} M. Heusler, Liv. Rev. Rel. {\bf 1}, 6 (1998).

\bibitem{zet}
O. Obregon, H. Quevedo, and M. P. Ryan, Time and ``angular" dependent
backgrounds from stationary axisymmetric solutions,
Phys. Rev. D {\bf 68}, (2004) in press, [gr-qc/0404003].



\bibitem{samos} H. Quevedo and M. Ryan, in
{\it Mathematical and Quantum Aspects of
Relativity and Cosmology}, edited by  S. Cotsakis and G.W. Gibbons
(Springer-Verlag, Berlin, 2000), [gr-qc/0305001].




\end{thebibliography}
\end{document}